\documentclass{emulateapj}

\shorttitle{Same AGN Fraction in Clumpy and Smooth Galaxies at $z \sim 2$}
\shortauthors{Trump et al.}

\begin{document}

\title{No More Active Galactic Nuclei in Clumpy Disks Than in Smooth
  Galaxies at $z \sim 2$ in CANDELS / 3D-HST\altaffilmark{*}}

\author{
  Jonathan R. Trump,\altaffilmark{1,$\dagger$}
  Guillermo Barro,\altaffilmark{2}
  St\'{e}phanie Juneau,\altaffilmark{3}
  Benjamin J. Weiner,\altaffilmark{4}
  Bin Luo,\altaffilmark{1}
  Gabriel B. Brammer,\altaffilmark{5}
  Eric F. Bell,\altaffilmark{6}
  W. Niel Brandt,\altaffilmark{1}
  Avishai Dekel,\altaffilmark{7}
  Yicheng Guo,\altaffilmark{2}
  Philip F. Hopkins,\altaffilmark{8}
  David C. Koo,\altaffilmark{2}
  Dale D. Kocevski,\altaffilmark{9}
  Daniel H. McIntosh,\altaffilmark{10}
  Ivelina Momcheva,\altaffilmark{11}
  S. M. Faber,\altaffilmark{2}
  Henry C. Ferguson,\altaffilmark{5}
  Norman A. Grogin,\altaffilmark{5}
  Jeyhan Kartaltepe,\altaffilmark{5}
  Anton M. Koekemoer,\altaffilmark{5}
  Jennifer Lotz,\altaffilmark{5}
  Michael Maseda,\altaffilmark{12}
  Mark Mozena,\altaffilmark{2}
  Kirpal Nandra,\altaffilmark{13}
  David J. Rosario,\altaffilmark{13}
  and Gregory R. Zeimann\altaffilmark{1}
}

\altaffiltext{*}{
  Based on observations with the NASA/ESA \emph{Hubble Space
  Telescope}, obtained at the Space Telescope Science Institute, which
  is operated by AURA Inc, under NASA contract NAS 5-26555.
\label{candelsobs}}

\altaffiltext{1}{
  Department of Astronomy and Astrophysics, 525 Davey Lab, The
  Pennsylvania State University, University Park, PA 16802, USA
\label{PSU}}

\altaffiltext{2}{
  University of California Observatories/Lick Observatory and
  Department of Astronomy and Astrophysics, University of California,
  Santa Cruz, CA 95064, USA
\label{UCO/Lick}}

\altaffiltext{3}{
  Irfu/Service d'Astrophysique, CEA-Saclay, Orme des Merisiers, 91191
  Gif-sur-Yvette Cedex, France
\label{Saclay}}

\altaffiltext{4}{
  Steward Observatory, University of Arizona, 933 North Cherry Avenue,
  Tucson, AZ 85721, USA
\label{Arizona}}

\altaffiltext{5}{
  Space Telescope Science Institute, 3700 San Martin Drive, Baltimore,
  MD 21218, USA
\label{STScI}}

\altaffiltext{6}{
  Department of Astronomy, University of Michigan, 500 Church St., Ann
  Arbor, MI 48109, USA
\label{Michigan}}

\altaffiltext{7}{
  Center for Astrophysics and Planetary Science, Racah Institute of
  Physics, The Hebrew University, Jerusalem 91904, Israel
\label{Jerusalem}}

\altaffiltext{8}{
  California Institute of Technology, MC 105-24, 1200 East California
  Boulevard, Pasadena, CA 91125 USA
\label{Caltech}}

\altaffiltext{9}{
  Department of Physics and Astronomy, University of Kentucky,
  Lexington, KY 40506, USA
\label{Kentucky}}

\altaffiltext{10}{
  Department of Physics and Astronomy, University of Missouri-Kansas
  City, 5110 Rockhill Rd, Kansas City, MO 64110, USA
\label{Missouri}}

\altaffiltext{11}{
  Department of Astronomy, Yale University, New Haven, CT 06520, USA
\label{Yale}}

\altaffiltext{12}{
  Max-Planck-Institut f\"{u}r Astronomie (MPIA), K\"{o}nigstuhl 17,
  D-69117 Heidelberg, Germany
\label{MPIA}}

\altaffiltext{13}{
  Max-Planck-Institut f\"{u}r extraterrestrische Physik (MPE),
  Giessenbachstrasse 1, D-85748 Garching bei M\"{u}nchen, Germany
\label{MPE}}

\altaffiltext{$\dagger$}{
  Hubble Fellow
\label{HF}}

\def\etal{et al.}
\newcommand{\Ha}{\hbox{{\rm H}$\alpha$}}
\newcommand{\Hb}{\hbox{{\rm H}$\beta$}}
\newcommand{\OII}{\hbox{[{\rm O}\kern 0.1em{\sc ii}]}}
\newcommand{\NeIII}{\hbox{[{\rm Ne}\kern 0.1em{\sc iii}]}}
\newcommand{\OIII}{\hbox{[{\rm O}\kern 0.1em{\sc iii}]}}
\newcommand{\NII}{\hbox{[{\rm N}\kern 0.1em{\sc ii}]}}
\newcommand{\SII}{\hbox{[{\rm S}\kern 0.1em{\sc ii}]}}
\newcommand{\HII}{\hbox{{\rm H}\kern 0.1em{\sc ii}}}

\begin{abstract}

  We use CANDELS imaging, 3D-HST spectroscopy, and Chandra X-ray data
  to investigate if active galactic nuclei (AGNs) are preferentially
  fueled by violent disk instabilities funneling gas into galaxy
  centers at $1.3<z<2.4$.  We select galaxies undergoing gravitational
  instabilities using the number of clumps and degree of patchiness as
  proxies.  The CANDELS visual classification system is used to
  identify 44 clumpy disk galaxies, along with mass-matched comparison
  samples of smooth and intermediate morphology galaxies.  We note
  that, despite being being mass-matched and having similar star
  formation rates, the smoother galaxies tend to be smaller disks with
  more prominent bulges compared to the clumpy galaxies.  The lack of
  smooth extended disks is probably a general feature of the $z \sim
  2$ galaxy population, and means we cannot directly compare with the
  clumpy and smooth extended disks observed at lower redshift.  We
  find that $z \sim 2$ clumpy galaxies have slightly enhanced AGN
  fractions selected by integrated line ratios (in the mass-excitation
  method), but the spatially resolved line ratios indicate this is
  likely due to extended phenomena rather than nuclear AGNs.
  Meanwhile the X-ray data show that clumpy, smooth, and intermediate
  galaxies have nearly indistinguishable AGN fractions derived from
  both individual detections and stacked non-detections.  The data
  demonstrate that AGN fueling modes at $z \sim 1.85$ - whether
  violent disk instabilities or secular processes - are as efficient
  in smooth galaxies as they are in clumpy galaxies.

\end{abstract}


\section{Introduction}

The observed correlations between the mass of a supermassive black
hole (SMBH) and the properties of its host galaxy bulge
\citep[e.g.][]{mag98,gul09} imply that SMBH and bulge growth may be
linked.  To maintain the SMBH-bulge relations over a galaxy's long
evolutionary history, rapid SMBH accretion in active galactic nucleus
(AGN) phases probably coincide with periods of high star formation
rate in the host galaxy.  Indeed, rapidly accreting AGNs are observed
to be predominantly located in the rapidly star-forming galaxies, from
the local universe \citep{kau03b,tru13b} to higher redshifts
\citep{mul12,har12,chen13,ros13a,ros13b}.  But the detailed physical
processes behind the coupled growth of SMBHs and galaxies remain
mysterious.


A key difficulty for any AGN/galaxy coevolution model is in
efficiently funneling gas down to the SMBH sphere of influence.
Gas-rich major mergers of galaxies are an effective means to
accomplish this, simultaneously (or near-simultaneously) igniting both
a starburst and a luminous AGN \citep{san88,dim05,hop06}.  However
many observations indicate that AGNs do not prefer merger remnant
hosts \citep{gro05,gab09, cis11, koc12}, and major mergers are likely
to fuel only nearby AGNs \citep{koss10,ell11} or the rare population
of very luminous quasars \citep{tru11c,tre12}.

\begin{figure*}[t] 
\begin{center}
\epsscale{1.1}
{\plotone{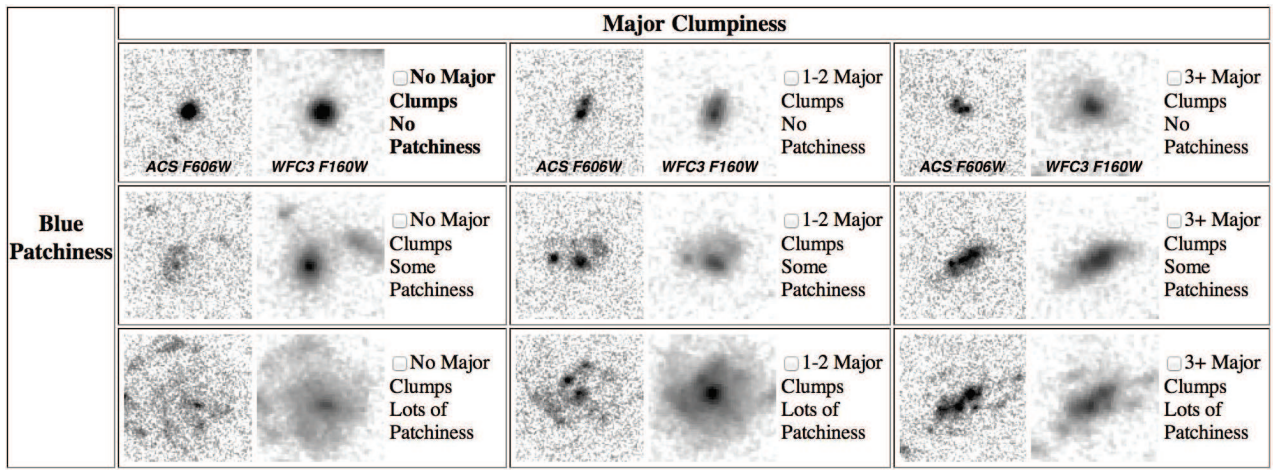}}
\end{center}
\figcaption{The diagram used for visual classification of clumpiness.
  The classification grid includes 9 template galaxies, each with
  images in the ACS F606W (rest-frame UV) and WFC3 F160W (rest-frame
  optical).  For each galaxy in GOODS-S, classifiers chose the
  template galaxy which best matched in both bands (with a focus on
  the F606W, where clumps are most visible).  We sum both axes and
  average over the 5 classifiers to assign each galaxy a ``total
  clumpiness'' $C$ from 0 (top left) to 4 (bottom right), motivated by
  theoretical simulations which demonstrate that patchiness might
  simply be clumpiness reprocessed by dust.  Clumpy galaxies have $C
  \geq 2$, smooth galaxies have $C \leq 1$, and intermediate galaxies
  have $1<C<2$.
\label{fig:classification}}
\end{figure*}

Violent disk instabilities might provide another solution to the gas
inflow problem \citep{bou11}.  These models are supported by the
observation that massive galaxies at $z \sim 2$ frequently have most
of their star formation in massive ($10^8$-$10^9$ $M_\odot$,
$100-500$~pc) clumps within irregular galaxies
\citep{cow95,elm04,rav06,guo12}, unlike the much smoother morphologies
of similar-mass star-forming galaxies in the local universe.  The
kinematics of these clumps indicate that they are not accreted as
minor mergers, but instead form as in-situ gravitational instabilities
\citep{sha08,bou08,for09,gen11,man14}.  The high gas fractions
observed in $z \sim 2$ galaxies \citep{tac10,dad10}, likely accreted
as cold gas along filaments of the cosmic web
\citep{ker05,dek06,dek09a}, would naturally result in highly turbulent
star-forming clumps \citep{her89,shlos93,dek09b}.

Violent disk instabilities might lead to clumps rapidly falling into
the center of a galaxy and efficiently fueling a luminous AGN
\citep{bou11}.  This remains a point of debate, however, as most
simulations lack the resolution and detailed physics to follow the gas
all the way into a galaxy's center.  Observations indicate that
star-forming clumps at $z \sim 2$ typically make up only a small
fraction of their galaxy's stellar mass, suggesting that they may be
too short-lived to migrate all the way into a galaxy center
\citep{wuyts12}.  Indeed, simulations show that clumps might be
destroyed by feedback \citep{hop12,hop13} or exhausted by star
formation \citep{for14} before reaching the nuclear AGN.  On the other
hand, other simulations find that clumps are sufficiently long-lived
to form central bulges and fuel AGNs even with feedback and mass loss
\citep{bou14}, and the gas between clumps, rather than the clumps
themselves, may be more important for large-scale inflows anyway
\citep[e.g.][]{dek13}.  In support of AGN fueling from violent disk
instabilities, \citet{bou12} found direct observational evidence for a
higher AGN fraction in a small sample (14) of $z \sim 0.7$ clumpy
galaxies compared to smooth (non-clumpy) extended disks matched in
stellar mass and redshift.

Here we extend the search for a connection between clumps and AGNs to
$1.3<z<2.4$, using data from the Cosmic Assembly Near-Infrared Legacy
Survey \citep[CANDELS][]{candels,candels2}.  This redshift range is
particularly interesting because it represents the peak of both cosmic
star formation and the AGN luminosity function
\citep[e.g.][]{cosmicsfr,aird10}.  Cold gas inflows and resultant
violent disk instabilities are also most prominent in simulations at
$z \sim 2$ \citep{ker05,dek09b}.  And in contrast to the local
universe, AGNs at $z>1$ may be unique in growing before a host galaxy
bulge develops, with evidence from both evolution in BH-bulge
relations \citep{peng06,jahnke09,ben11,cis11} and from stacked
detection of AGN signatures in low-mass disk galaxies
\citep{tru11b,xue12}.  The gravitational torques provided by violent
disk instabilities provide a potential way to fuel an AGN in a disk
galaxy even while the bulge is still under construction.

Section 2 describes the combination of Hubble Space Telescope ({\it
  HST}) Wide Field Camera 3 (WFC3) near-infrared (IR) imaging and
spectroscopy used to construct and study our samples of $z \sim 2$
galaxies with clumpy, smooth, and intermediate morphologies.  We
identify AGNs using X-ray emission, emission-line diagnostics, and
spatially resolved emission-line ratios: as Section 3 describes, this
includes a broad range of AGN properties and avoids contamination by
shock-dominated galaxies.  We put these selection techniques together
in Section 4, and demonstrate that clumpy galaxies are not more likely
to host AGNs than smoother galaxies.  Section 5 concludes with a
discussion of what these results mean for AGN/galaxy coevolution at $z
\sim 2$.  Throughout the paper we use a standard $\Lambda$CDM
cosmology with $h_0=0.7$.

\begin{deluxetable*}{ccrrrrrr}
  \tablecolumns{8}
  \tablecaption{Galaxy Properties\label{tbl:catalog}}
  \tablehead{
    \colhead{ID} & 
    \colhead{Morphology} & 
    \colhead{RA} & 
    \colhead{Dec} & 
    \colhead{$z$} & 
    \colhead{$f(\Hb)$} & 
    \colhead{$f(\OIII)$} & 
    \colhead{$\log(M_*)$} \\
    \colhead{-} & 
    \colhead{-} & 
    \colhead{(deg)} & 
    \colhead{(J2000)} & 
    \colhead{-} & 
    \colhead{$10^{-18}$~erg~s$^{-1}$~cm$^{-2}$} & 
    \colhead{$10^{-18}$~erg~s$^{-1}$~cm$^{-2}$} & 
    \colhead{$(\log(M_{\odot}))$} }
  \startdata
 7502 &  clumpy  &  53.07221603 & -27.84033012 &  1.61 & $99.95\pm28.56$ & $83.47\pm24.51$ &  9.63 \\
 7897 &  clumpy  &  53.14831924 & -27.83686638 &  1.90 & $23.49\pm12.22$ & $27.09\pm8.42$ &  9.82 \\
 7907 &  clumpy  &  53.15573502 & -27.83712769 &  1.39 & $18.95\pm13.73$ & $169.79\pm21.69$ &  9.67 \\
 7930 &  clumpy  &  53.15451050 & -27.83648491 &  2.03 & $14.20\pm10.43$ & $43.64\pm10.26$ &  9.41 \\
 8206\tablenotemark{a} &  clumpy  &  53.14358902 & -27.83471107 &  1.99 & $10.82\pm5.73$ & $39.97\pm10.54$ &  9.85 \\
 5023 &  smooth  &  53.10656738 & -27.86481476 &  1.90 & $<3.31$ & $18.62\pm5.92$ &  9.74 \\
 5728 &  smooth  &  53.08922577 & -27.85734558 &  2.03 & $16.80\pm7.95$ & $79.73\pm11.31$ & 10.37 \\
 6278\tablenotemark{b} &  smooth  &  53.06018448 & -27.85304642 &  1.54 & $<2.73$ & $240.61\pm21.63$ & 10.79 \\
 6678 &  smooth  &  53.07507324 & -27.84802055 &  1.73 & $37.84\pm7.38$ & $21.91\pm6.77$ & 10.14 \\
 7806 &  smooth  &  53.16970062 & -27.83803558 &  1.94 & $8.18\pm7.71$ & $38.97\pm11.10$ & 10.37 \\
 5552 &  intermed  &  53.06779480 & -27.85925293 &  1.55 & $15.67\pm14.78$ & $73.40\pm16.57$ &  9.81 \\
 6895 &  intermed  &  53.09495926 & -27.84582710 &  1.56 & $<4.50$ & $49.29\pm8.75$ &  9.44 \\
 7952 &  intermed  &  53.18961334 & -27.83626175 &  2.09 & $39.79\pm13.92$ & $52.87\pm9.57$ &  9.89 \\
 8124 &  intermed  &  53.10293198 & -27.83484268 &  1.38 & $55.60\pm21.62$ & $102.88\pm15.12$ &  9.18 \\
 8307 &  intermed  &  53.07443237 & -27.83325577 &  1.54 & $<6.95$ & $55.32\pm10.17$ &  9.26  \enddata
 \tablecomments{Weak \Hb\ lines detected below a 1$\sigma$ threshold
   are treated as upper limits (using the 1$\sigma$ error as the
   limit).  The full catalog of 44 clumpy galaxies, 41 smooth
   galaxies, and 35 intermediate galaxies appears as a
   machine-readable table in the electronic version.}
 \tablenotetext{a}{X-ray detected, but in the soft band only and
   consistent with emission from a star-forming galaxy \citep{xue11}.}
 \tablenotetext{b}{X-ray detected and classified as an AGN by
   \citet{xue11}.}
\end{deluxetable*}

\section{Observational Data}

We select a sample of 44 clumpy galaxies from the Great Observatories
Origins Deep Survey South \citep[GOODS-S][]{goods} region of CANDELS.
For comparison, we also construct mass-matched samples of 41 smooth
(non-clumpy) and 35 intermediate galaxies.  All galaxies have $H<24$
(to ensure reliable classification of clumpiness) and have $\OIII$
detected at the 3$\sigma$ level (for reliable AGN line ratio
diagnostics) in the redshift range $1.3<z<2.4$.  The effective limits
on star formation rate (SFR) and stellar mass ($M_*$) caused by these
flux limits are discussed in Section 2.3.  Each morphology category
has a median redshift of $z=1.85$.  Redshifts and emission line
measurements come from {\it HST}/WFC3 grism spectroscopy taken by the
3D-HST survey \citep{3dhst}.

The observational data are described below, with particular attention
to the methods for clumpiness classification and AGN identification.
The derived data for the clumpy, smooth, and intermediate galaxies are
presented in Table \ref{tbl:catalog}.

\subsection{Visual Morphologies: Clumpy, Smooth, and Intermediate}

\begin{figure}[t] 
\epsscale{1.15}
{\plotone{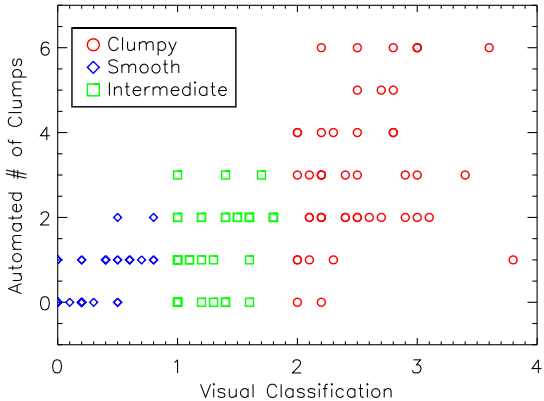}}
\figcaption{Comparison between the automated number of clumps from
  \citet{guo14} and the visual clumpiness used in this work.  There is
  a loose correlation between the two measures of clumpiness, matching
  the $\sim$75\% agreement between automated and visual methods found
  by \citet{guo14}.
\label{fig:clumpcompare}}
\end{figure}

Our samples of clumpy, smooth, and intermediate galaxies come from the
4-epoch CANDELS GOODS-S visual classification catalog \citep{kar14}.
Each galaxy has high-resolution {\it HST} imaging in the rest-frame UV
from ACS F606W and F850LP and in the rest-frame optical from WFC3
F125W and F160W \citep{candels,candels2}.  Galaxies were classified on
the clumpiness/patchiness grid shown in Figure
\ref{fig:classification}.  Here ``clumpiness'' is a measure of the
number of compact knots, while ``patchiness'' refers to the more
diffuse irregularities within galaxies.

We use visual classification because automated selection of clumps and
patchiness remains a difficult computational problem
\citep[e.g.][]{guo12}.  Inspectors were instructed to focus on the
bluest passband for clumpiness classification, as clumps are typically
most evident in in the rest-UV \citep{guo12}.  All galaxies in our
sample (from the 4-epoch catalog of \citealt{kar14}) were inspected by
at least five classifiers, and we combine the different
classifications into a single averaged result for each galaxy.

\begin{figure*}[t] 
\epsscale{1.15}
{\plotone{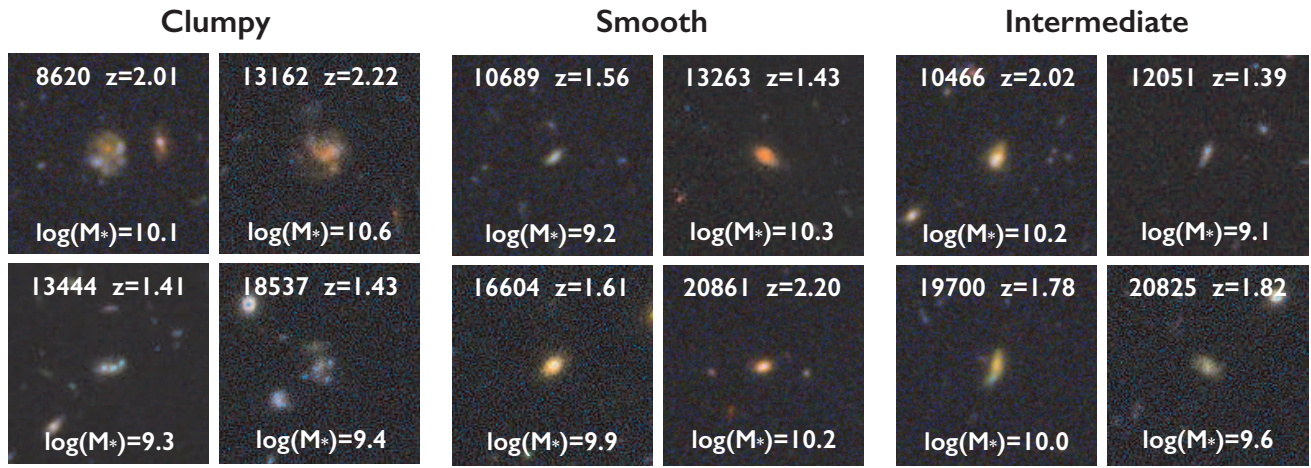}}
\figcaption{Images of four clumpy, four smooth, and four
  intermediate-morphology galaxies, chosen to be representative in
  redshift and stellar mass (stellar masses given in each panel are in
  units of solar mass).  Each thumbnail is 10$\arcsec$ on a side and
  is a color-composite using the {\it HST}/ACS $i$ and WFC3 $JH$ band
  observations.
\label{fig:images}}
\end{figure*}

We use both clumpiness and patchiness as indicators of violent disk
instabilities within a galaxy.  This is motivated by the presence of
both compact/round clumps and diffuse/elongated structures in
simulations of $z \sim 2$ galaxies undergoing violent disk
instabilities \citep{cev12,man14}.  In addition, if star-forming
clumps are enshrouded by dust, their emission would likely be
reprocessed in a patchy morphology.  Observations show that galaxies
appearing clumpy in the UV are merely patchy morphologies in the
near-infrared: adding dust to such a system would cause it to appear
patchy in the UV rather than clumpy.  We also note that \citet{guo14}
find better agreement between automated clump-finding and visual
classification when patchiness is included in the visual clumpiness.
Therefore we combine both axes in Figure \ref{fig:classification} for
the identification of clumpy galaxies.


The classifications on each axis are summed to give each galaxy a
clumpiness+patchiness value ranging from 0 to 4.  Here 0 represents
the top left (no clumpiness, no patchiness) and 4 represents the lower
right (3+ clumps and maximally patchy).  The individual classification
results are averaged together to assign ``total clumpiness'' parameter
$C$, which we use to define the galaxies most and least likely to be
dominated by violent disk instabilities.  We define $C \geq 2$
galaxies as ``clumpy,'' $C<1$ galaxies as ``smooth,'' and $1 \leq C<2$
galaxies as ``intermediate.''  These divisions are chosen such that
each category has roughly the same number of galaxies, for ease of
comparing AGN fractions across morphologies.

We compare the visual classification $C$ with the automated number of
clumps from \citep{guo14} in Figure \ref{fig:clumpcompare}.  There is
a loose correlation between the two methods: the visually-identified
smooth galaxies tend to have the fewest automated number of clumps,
while the visually clumpy galaxies typically have the most automated
number of clumps.  There is some scatter due to the imperfection of
both methods.  \citet{guo14} include a more detailed comparison of
automated clump-finding with the same visual classifications used
here, finding $\sim$75\% agreement between the two methods (see their
Appendix).

Figure \ref{fig:images} shows {\it HST} $iJH$ (rest UV-optical) color
composite images for several representative clumpy, smooth, and
intermediate galaxies in our sample.  Images of all galaxies are
additionally shown in the Appendix (Figures \ref{fig:clumpyimages},
\ref{fig:smoothimages}, and \ref{fig:intimages}).

\subsection{HST/WFC3 G141 Slitless Grism}

The GOODS-S region of CANDELS has near-complete spectroscopic coverage
in the near-IR from publicly available 2-orbit {\it HST}/WFC3 G141
grism observations taken by the 3D-HST survey \citep{3dhst}.  For
redshifts and line measurements of individual galaxies, the data were
reduced using the {\tt aXe} software \citep[][available at {\tt
  http://axe.stsci.edu/axe/}]{kum09} to produce 2D and 1D wavelength-
and flux-calibrated spectra at $1.1<\lambda<1.7\mu$m.  We used the
{\tt specpro} IDL software\footnote{The specpro package is available
  at {\tt http://specpro.caltech.edu}} \citep{specpro} to visually
inspect and determine cross-correlation redshifts for the samples of
clumpy, smooth, and intermediate galaxies.  Spectra with significant
($>$10\%) contamination (from neighboring objects) in the \Hb+\OIII\
region are rejected from each sample: this occurs for about $\sim$15\%
of galaxies.

\begin{figure}[t] 
\epsscale{1.1}
{\plotone{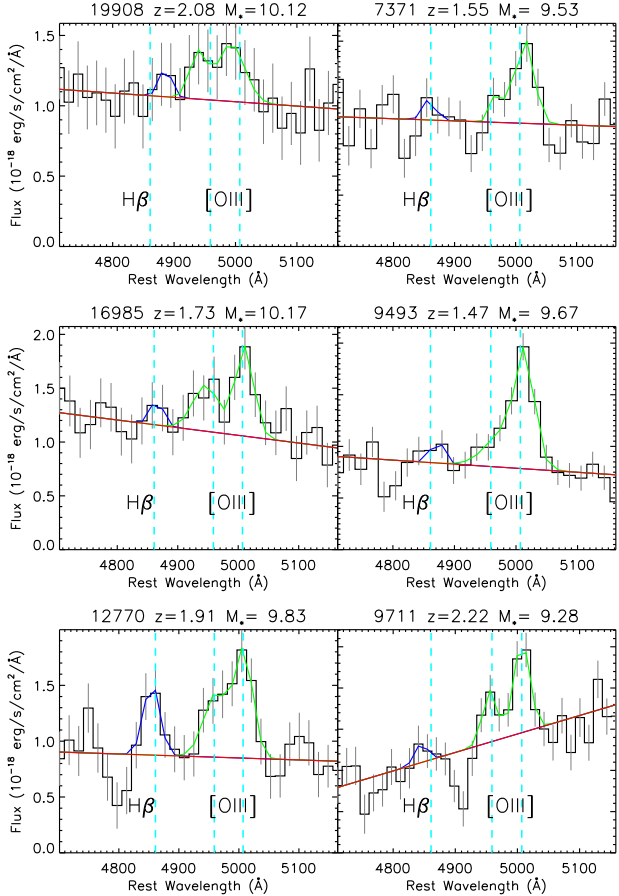}}
\figcaption{Example line fits to the \Hb+\OIII\ regions.  The blended
  lines are simultaneously fit by three Gaussians, with the \Hb\ flux
  measured directly from the \Hb\ Gaussian fit, and the
  \OIII$\lambda$5007 line computed as $3/4$ of the two \OIII\
  Gaussians.  The top row shows spectra of two clumpy galaxies, the
  middle row is smooth galaxies, and the bottom row is intermediate
  galaxies.  Both 19908 and 16985 have weak \Hb\ emission detected
  below 1$\sigma$ significance, and for these galaxies the 1$\sigma$
  errors in \Hb\ are treated as upper limits.
\label{fig:linefits}}
\end{figure}

Line fluxes and ratios of \Hb\ and \OIII$\lambda$5007\AA\ were also
measured from the {\tt aXe}-reduced 3D-HST spectra.  The low
resolution of the WFC3 slitless grism ($R \simeq 130$ and
46.5\AA/pixel, worse for extended sources) means that the \Hb\ and
\OIII$\lambda$4959,5007\AA\ lines are somewhat blended.  We
simultaneously fit all three lines with Gaussians, constraining each
line to be fit within (rest-frame) 50\AA\ ($\sim$2-3 pixels) of the
line center.  We do not constrain the width of each line because the
spatial broadening of the slitless grism frequently causes unusual
profiles which can differ for each line.  The \Hb\ line flux is
measured directly from the Gaussian fit, while the \OIII$\lambda$5007
flux is measured as $3/4$ of the total flux from both blended \OIII\
Gaussians \citep{sto00}.  If we instead fix the two $\OIII$ Gaussians
to have total fluxes with a 1:3 ratio, the measured line fluxes do not
significantly change.  Examples of the line fits are shown in Figure
\ref{fig:linefits}.

Uncertainties in the line ratios are computed by re-fitting continua
and Gaussians on 10000 realizations of the resampled data.  Our sample
includes only galaxies with \OIII$\lambda$5007\AA\ fluxes measured at
the 3$\sigma$ level.  We do not set any requirement on the $\Hb$ line
flux measurements, and $\Hb$ line fluxes less than the 1$\sigma$ error
are treated as upper limits (using the 1$\sigma$ error as the limit).
In general, the unique asymmetric shape of the blended $\OIII$ lines
enables secure redshifts even when $\Hb$ is poorly detected.

To study spatially resolved line ratios, we used separate reductions
of the WFC3 G141 data from the 3D-HST pipeline, as described in
\citet{3dhst,bra13}.  The 3D-HST reduction method produces superior
spatially resolved spectra because it interlaces rather than drizzles:
this mitigates the correlated noise associated with drizzling in the
final high-resolution ($0\farcs06$/pixel) combined 2D spectra.

\begin{figure}[t]
\epsscale{1.15}
{\plotone{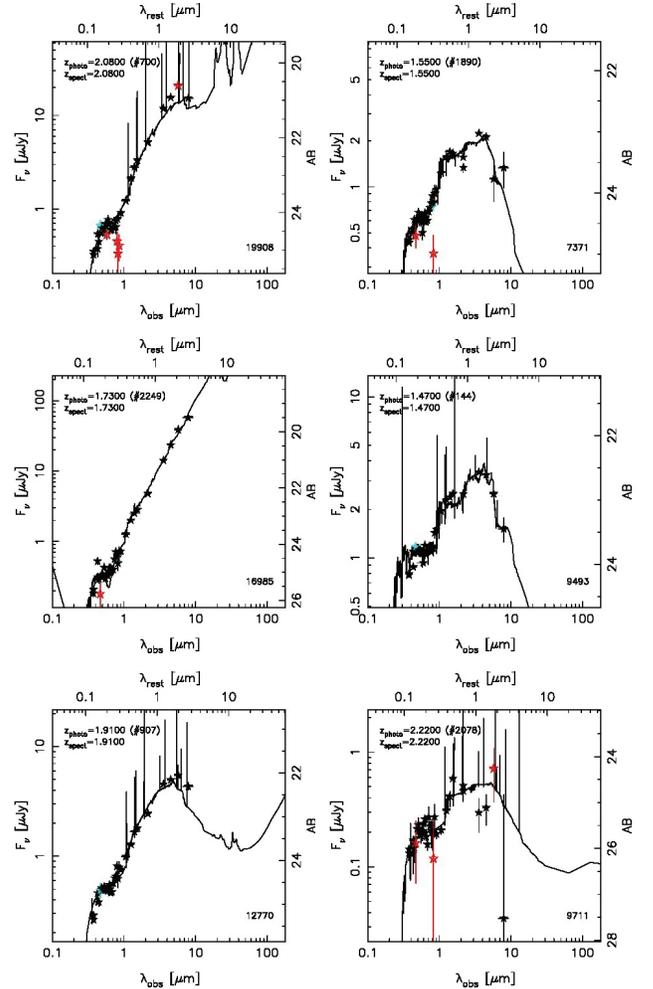}}
\figcaption{Best-fit models to the SEDs of the same galaxies shown in
  Figure \ref{fig:linefits}.  Stellar masses are computed from these
  best-fit \citet{bru03} models.  The SED of the galaxy 16985 (which
  is an X-ray AGN) has red near-IR colors suggestive of an AGN
  \citep[e.g.][]{don12}: it is the only galaxy in the sample which
  might have significant AGN contribution to the SED.
\label{fig:sedfits}}
\end{figure}

\subsection{Stellar Masses and Star Formation Rates}

We calculate stellar masses and star formation rates for the clumpy,
smooth, and intermediate galaxies using the extensive UV / optical /
IR photometry in GOODS-S, with 18 bands including 8 with
high-resolution {\it HST} imaging \citep{guo13}.  First, the
spectroscopic redshift from the {\it HST}/WFC3 grism is used to shift
the observed photometry to the rest-frame.  We then fit the rest-frame
spectral energy distribution (SED) with \citet{bru03} models that
include a \citet{cha03} initial mass function, exponentially declining
star-formation histories, and a \citet{cal01} extinction law.  Stellar
mass is given by the best-fit model.  Example SED fits (for the same
galaxies shown in Figure \ref{fig:linefits} are shown in Figure
\ref{fig:sedfits}.

\begin{figure}[t]
\epsscale{1.15}
{\plotone{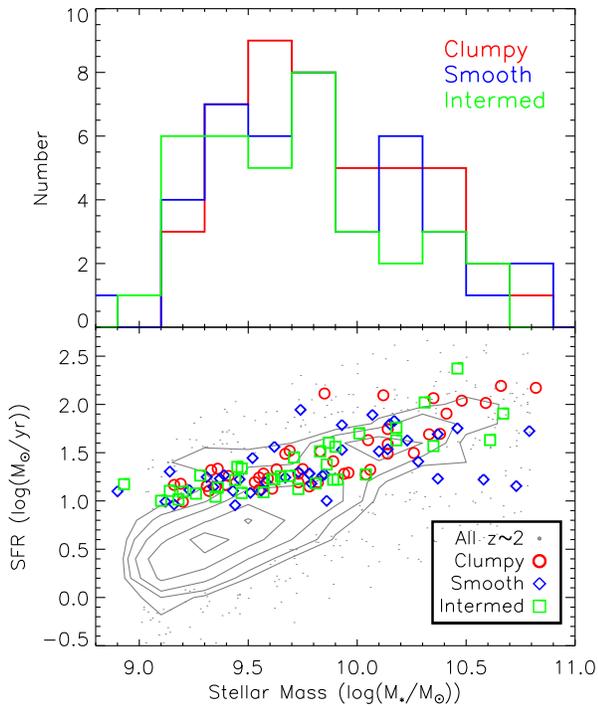}}
\figcaption{{\it Top:} Histograms showing stellar mass distribution
  for each sample.  The smooth and intermediate galaxies are chosen to
  have the same mass distribution as the clumpy galaxies.  {\it
    Bottom:} Star formation rate vs. stellar mass for the clumpy,
  smooth, and intermediate galaxies studied in this work (colored
  points).  The contours show the larger population of galaxies with
  $1.3<z_{\rm phot}<2.4$ in the same GOODS-S region.  While the SFR is
  estimated from the \citet{wuyts11} UV+IR method, it is correlated
  with the $\OIII$ emission line flux, and the observed SFR$\gtrsim
  10$~M$_\odot$~yr$^{-1}$ limit is a consequence of the 3$\sigma$
  \OIII\ detection limit.  Meanwhile the $H<24$ selection causes the
  $M_* \gtrsim 10^9M_\odot$ limit.
\label{fig:sfrmass}}
\end{figure}

There is some evidence that $z \sim 2$ galaxies are more likely to
have constant star-formation histories rather than the exponentially
declining models used here.  In practice, our best-fit SEDs generally
have $\tau$ values which are very long (parameterizing the star
formation history as $SFR \sim e^{-t/\tau}$).  We also tested the
effects of forcing a constant star formation history, and found the
masses to change by only $\lesssim$0.1~dex in all cases.

Meanwhile star formation rates (SFRs) are calculated following the
method of \citet{wuyts11} when mid- and far-IR data are available, and
SED-fitting following the method of \citet{bar13} otherwise.  IR
emission from an intrinsically luminous AGN might contaminate the IR
with a hot dust bump \citep[effectively observed as red IRAC colors,
e.g.][]{don12}, influencing both the stellar mass and SFR estimates.
This occurs for one of our objects: the clumpy galaxy and X-ray AGN
16985, shown in Figure \ref{fig:sedfits}.  The remainder of the sample
have well-fit galaxy SEDs without evidence for significant AGN
contribution.  This is unsurprising, as only intrinsically luminous
($log(L_X) \gtrsim 44$) AGNs tend to have hot dust IR emission which
dominates over the galaxy emission \citep{tru11a,don12}.

\begin{figure*}[t] 
\epsscale{1.15}
{\plotone{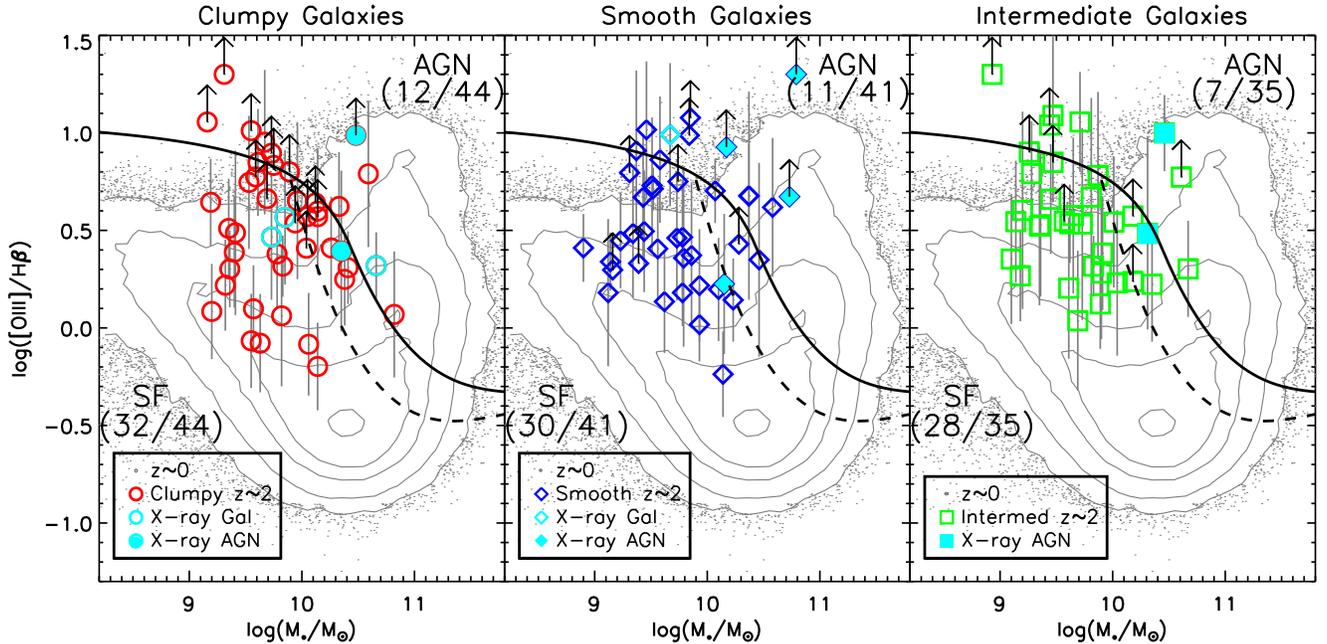}}
\figcaption{The mass-excitation (MEx) diagnostic \citep{jun11} of
  $\OIII/\Hb$ line ratio versus stellar mass.  Panels from left to
  right show clumpy galaxies, smooth galaxies, and
  intermediate-morphology galaxies.  Cyan symbols indicate galaxies
  detected in X-rays: filled symbols are X-ray AGN, while open cyan
  symbols are consistent with X-ray galaxies (i.e., with X-rays
  powered by non-AGN processes).  For comparison, each panel also
  shows a $z \sim 0$ galaxy sample from the SDSS.  The fraction of MEx
  AGN is slightly higher for the clumpy galaxies (29\% AGN) than for
  the samples of smooth (27\% AGN) and intermediate (20\% AGN)
  galaxies.  However we show in Sections 3.2 and 3.3 that this is
  likely due to extended high-ionization phenomena (e.g., clumpy star
  formation) rather than nuclear AGN.
\label{fig:bpt}}
\end{figure*}

Figure \ref{fig:sfrmass} (top panel) shows the mass distributions for
each of the clumpy, smooth, and intermediate galaxy samples.  The
three samples are selected to be matched in stellar mass: starting
with the clumpy galaxy sample, we constructed the smooth and
intermediate galaxy samples to have a similar mass distribution.  All
three samples have median and mean $M_* = 10^{9.8}M_\odot$.

Our galaxies are plotted in $SFR-M_*$ space in Figure
\ref{fig:sfrmass} (bottom panel).  For comparison we also show the SFR
and $M_*$ of the larger population of GOODS-S galaxies with
photometric redshifts calculated by \citet{dah13} in the same
$1.3<z<2.4$ redshift range.  Our $z \sim 1.85$ galaxies are limited to
$M_* \gtrsim 10^9 M_\odot$ by the $H<24$ selection criterion.  The
requirement for \OIII\ detection at the 3$\sigma$ level imposes a flux
limit of $f(\OIII) \gtrsim 5 \times 10^{-17}$~erg~s$^{-1}$~cm$^{-2}$
in the 2-orbit WFC3 grism data.  Although we estimate SFR from the SED
\citep[following][]{wuyts11} rather than from emission lines, this
$\OIII$ line flux translates roughly to the observed SFR limit of
SFR$\gtrsim 10$~M$_\odot$~yr$^{-1}$ (assuming $f(\Ha) \sim f(\OIII)$
and using the \citealp{ken98} relation).  For the lower-mass half of
the sample ($M_*<10^{9.8}M_\odot$), this SFR limit restricts the
galaxies to be starbursting and above the star-forming ``main
sequence'' \citep[see also][]{noe07,dad07}.  On the other hand the
higher-mass half of the sample ($M_*>10^{9.8}M_\odot$) generally lies
on the main sequence for star-forming galaxies at $1.3<z<2.4$.  The
only notable exception is a few high-mass smooth galaxies below the
main sequence which might be quenching.

\subsection{X-ray Data}

The CANDELS GOODS-S field also contains 4~Ms of {\it Chandra} X-ray
data \citep{xue11}.  The 4~Ms depth corresponds to an on-axis flux
limit of $3.2 \times 10^{-17}$~erg~$s^{-1}$~cm$^{-2}$, which
corresponds to a luminosity limit of $L_X>10^{41.9}$~erg~s$^{-1}$ at
our sample's median redshift of $z=1.85$.  We use the classifications
of \citet{xue11} to separate X-ray galaxies (with X-ray detections
consistent with the $L_X-SFR$ relation, \citealt{leh10,min14}) from
the more luminous and hard-spectrum AGNs.  The X-ray data for
undetected sources in each morphology category are also stacked, as
discussed in Section 3.3.

\section{AGNs in Clumpy Galaxies?}

We compare the AGN fraction of clumpy galaxies with the AGN fraction
among the mass-matched smooth and intermediate galaxies.  AGN
detection is accomplished in three different ways: with standard line
ratio selection via the ``mass-excitation'' method \citep{jun11}, with
spatially resolved line ratios, and with X-rays (both for individual
sources and stacks of each morphological category).

\subsection{Mass Excitation Diagnostic for AGNs}

It has long been known that the high-ionization emission of AGNs
results in a different emission line signature than observed in
typical \HII\ regions associated with star formation (SF): in
particular, AGN narrow line regions tend to exhibit higher ratios of
collisionally excited ``forbidden'' lines to hydrogen recombination
lines \citep{seyf43,ost65}.  The classic \citet[][BPT]{bpt81} and
\citet[][VO87]{vo87} AGN/SF diagnostics use the ratios of
$f(\OIII\lambda5007)/f(\Hb)$ vs. $f(\NII\lambda6584)/f(\Ha)$ or
$f(\SII\lambda6718+6731)/f(\Ha)$: the small wavelength separation of
each line pair means that these ratios are effectively insensitive to
reddening.

The low resolution of the WFC3 grism means that the \NII\ and \Ha\
lines are not resolved from one another, and so the standard BPT or
VO87 diagnostics cannot be used on our data.  Instead we use the
``mass-excitation'' (MEx) method \citep{jun11}, which uses the
$\OIII/\Hb$ ratio with the stellar mass $M_*$ to separate AGN and SF
galaxies.  Functionally, the MEx method uses the correlation between
mass and metallicity \citep[e.g.][]{tre04} to separate low-metallicity
galaxies from AGN, both of which exhibit similarly high $\OIII/\Hb$
ratios.  Compared to X-ray or infrared surveys, line ratio selection
is often more sensitive to obscured and moderately accreting AGN, but
it also less reliable \citep[e.g.][]{jun14}.  There is also some
debate whether line ratio diagnostics remain applicable in $z>1$
galaxies at all, since higher redshift galaxies may have different gas
properties \citep{liu08,bri08,kew13}, changing their line ratios in
the absence of AGN \citep[but for counterarguments,
see][]{wri10,tru11b,tru13a}.  \citet{jun14} also demonstrate that the
different selection effects in high- and low-redshift samples have
important effects on their observed MEx distributions.  For this
reason we avoid comparing the $z \sim 1.5$ galaxy samples with local
objects.  Instead we simply compare clumpy, smooth, and intermediate
galaxies all within the same redshift range.

Figure \ref{fig:bpt} shows the MEx diagram for the clumpy, smooth, and
intermediate galaxies.  The solid line in each panel shows the
empirical line from \citet{jun14} dividing AGN and SF galaxies,
adjusted for the redshift ($z \sim 1.85$) and line flux limit
($f(\OIII) \gtrsim 5 \times 10^{-17}$~erg~s$^{-1}$~cm$^{-2}$):
\begin{equation}
  y = 0.375/(m-10.4)+1.14~{\rm if}~m \le 9.88
\end{equation}
\begin{equation}
  y = 290.2 - 76.34m + 6.69m^2 - 0.1955m^3~{\rm otherwise.}
\end{equation}
Here $y=\log(\OIII/\Hb)$ and $m = \log(M_*/M_\odot)-0.51$.  Assuming
that all galaxies above this line are AGNs, clumpy galaxies seem to
have the highest AGN fraction (12/44, 27\%), followed by smooth
galaxies (11/41, 27\%) and intermediate galaxies (7/35, 20\%).

However the simple binary classification of counting objects above and
below the AGN/SF line is not entirely appropriate, given that most
galaxies have emission lines with composite contribution from both
SMBH accretion and \HII\ regions.  The MEx diagram is much more suited
to a probabilistic classification approach, as introduced by
\citet{jun11}.  The MEx probabilities, updated for use at $z>1$ by
\citet{jun14} and assuming an error of 0.2~dex in $M_*$, indicate that
$39_{-6}^{+8}\%$ of clumpy, $36_{-6}^{+8}\%$ of smooth, and
$35_{-7}^{+9}\%$ of intermediate galaxies are AGN-dominated.  Given
the uncertainty of the MEx diagram at $z>1$
\citep[e.g.][]{kew13,new14,jun14}, these probabilities do not
necessarily represent absolute AGN fractions, but they remain useful
for comparing relative AGN fractions among the three morphology
classes.  The three AGN fractions are all consistent with one another,
and the clumpy galaxies have only marginally ($<$1$\sigma$) more AGNs
than smoother galaxies.  In the next two subsections we show that the
higher line ratios in clumpy galaxies are likely an effect of high
ionization in extended regions (due to shocks or dense star-forming
clumps) rather than nuclear AGN.

\subsection{Spatially Resolved Line Ratios}

The traditional BPT and MEx methods use line ratios integrated over
the entire galaxy, and thus may be diluted by star formation and/or
affected by non-nuclear ionization that has nothing to do with AGN
activity.  In particular, the violent disk instabilities of clumpy
galaxies may lead to high-velocity shocks or dense $\HII$ regions
which can mimic AGN-like line ratios \citep[e.g.][]{rich11,new14}.
For this reason we use the spatial resolution of the WFC3 grism
spectroscopy to go beyond integrated line ratios and investigate line
ratio gradients, following \citet{wri10,tru11b}.  Spatially resolved
line ratios can reveal if AGNs selected by integrated line ratios are
the product of nuclear AGNs or extended phenomena (like shocks or
high-ionization star formation).

\begin{figure}[t] 
\epsscale{1.15}
{\plotone{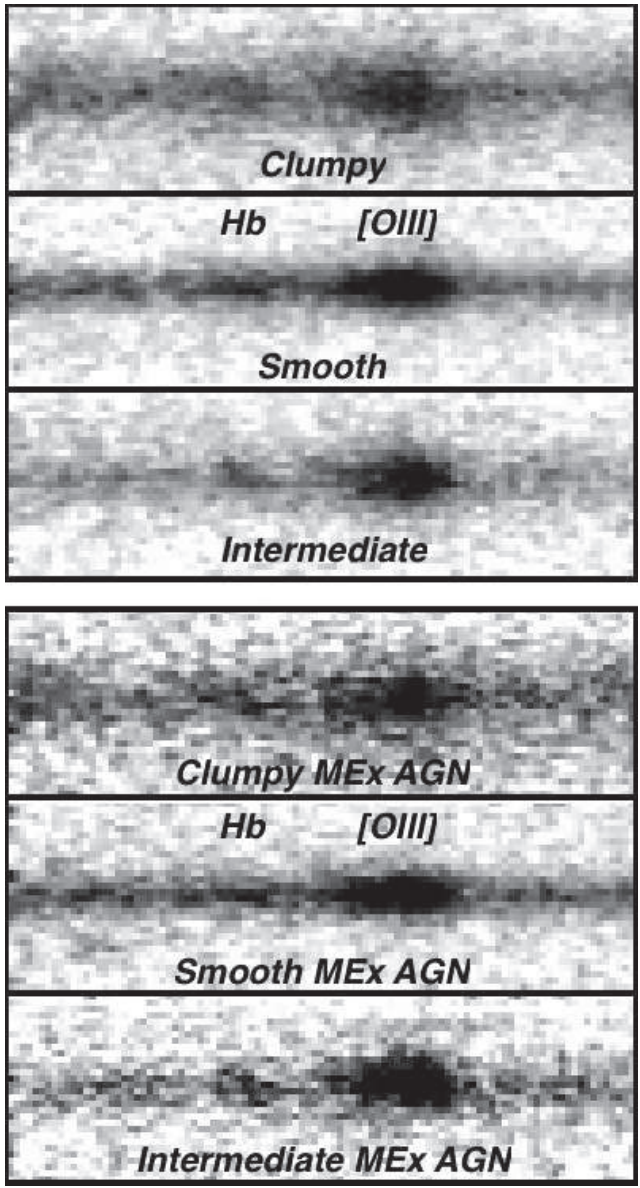}}
\figcaption{Stacked two-dimensional {\it HST}/WFC3 grism spectra for
  clumpy, smooth, and intermediate-morphology galaxies.  The top three
  spectra are stacks of all galaxies in each morphology class, while
  the bottom three are stacks of MEx AGNs only.  The emission line
  sizes in both the dispersion and cross-dispersion directions are a
  result of galaxy size rather than velocity due to the low resolution
  ($R \sim 300$) of the slitless grism.  The stacked data provide
  sufficient signal-to-noise to separately extract nuclear and
  extended one-dimensional spectra and line ratios.
\label{fig:stackimages}}
\end{figure}

The two-dimensional (2D) spectra of individual sources generally lack
the signal-to-noise for well-measured spatially resolved line ratios,
so we stack spectra of galaxies in each of the clumpy, smooth, and
intermediate galaxy samples.  Two stacks are constructed for each
morphology class: one with all galaxies, and another using only
galaxies classified as AGNs on the MEx diagram in Section 3.1.
Centers of galaxies are defined by the SExtractor \citep{sextractor}
coordinates in the F160W detection image, translated to a spectral
trace in the dispersed G141 2D spectrum using well-calibrated
polynomials from {\tt aXe} which are typically accurate to
$\lesssim$0.2~pixels.  The image centroid and resultant spectral trace
may not coincide with the brightest (continuum or emission line) flux
of a galaxy, particularly for clumpy galaxies: we investigate
centering on the brightest flux rather than image centroids in Section
4.1.  Each galaxy spectrum is also normalized by its total $\OIII$
flux so that the stack is equal-weighted and not dominated by the most
$\OIII$-luminous AGNs.  (This normalization means that the stacked
line ratios are slightly lower than the mean of the individual
ratios.)  The stacked 2D spectra of all galaxies in each morphology
class are shown in Figure \ref{fig:stackimages}.  It is immediately
evident that the clumpy galaxies are typically larger than the
intermediate or smooth galaxies: for a galaxy to be classified as
clumpy, it must be large enough for the clumps to be resolved in the
{\it HST} data.  Still, the smooth and intermediate galaxies are not
so small as to be unresolved, and they have enough signal beyond the
central three pixels to have well-measured extended (non-nuclear)
spectra.  We return to a discussion of the different sizes of the
clumpy and smooth galaxies in Section 4.2.

\begin{figure}[t] 
\epsscale{1.15}
{\plotone{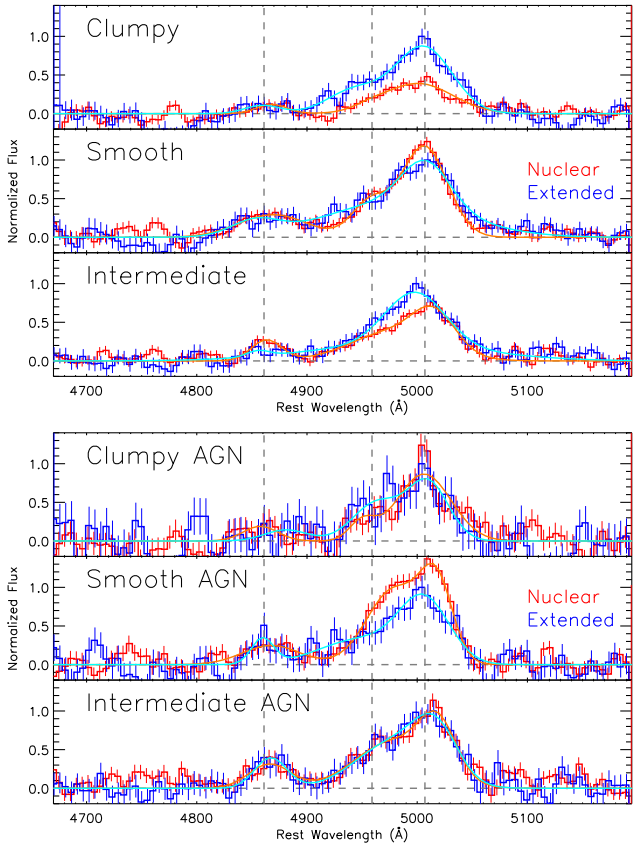}}
\figcaption{Nuclear and extended 1D spectra from the stacked {\it
    HST}/WFC3 grism data for clumpy, smooth, and
  intermediate-morphology galaxies.  The top three panels are the
  stacked spectra of all galaxies, while the bottom three are for
  galaxies classified as MEx AGNs.  The $\Hb$ and $\OIII$ line centers
  are shown by the dashed lines, and the orange and cyan lines give
  the best-fit Gaussians to the nuclear and extended 1D spectra.
\label{fig:stackspec}}
\end{figure}

We separately extract nuclear and extended one-dimensional (1D)
spectra from the stacked data, shown in Figure \ref{fig:stackspec}.
Here ``nuclear'' is defined as the central 3 pixels, corresponding to
$0\farcs18$ across and a $\sim$1~kpc radius at $1.3<z<2.4$.  The
``extended'' spectra are extracted from pixels 3-6 on either side of
the trace, translating to an annulus extending radially from
$\sim$2-4~kpc.  It is possible that some emission from the AGN narrow
line region extends into the extended aperture \citep{hai13,vdl13},
and the $\sim$1~kpc nuclear region is also likely to include some
galaxy starlight.  Although our regions will not perfectly disentangle
AGN and galaxy light, the nuclear region preferentially includes AGN
light and the extended region includes more galaxy starlight.  Thus
the spatially resolved line ratios are likely to be significantly more
sensitive than the integrated line ratios.

\begin{figure*}[ht] 
\epsscale{1.15}
{\plotone{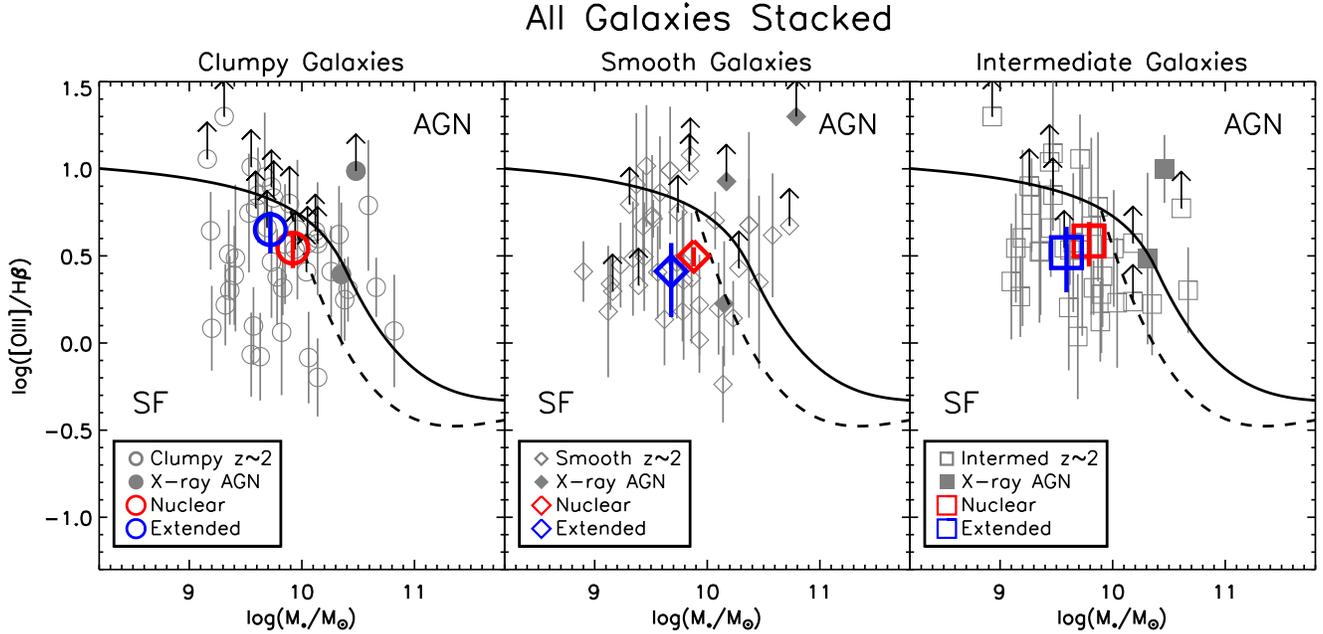}}
\figcaption{The nuclear and extended line ratios of stacked clumpy,
  smooth, and intermediate morphology galaxies.  These line ratios are
  placed in the MEx diagram using the median $M_*$ of each stack, with
  a small artificial offset between the nuclear and extended points.
  The integrated values for individual galaxies, seen in Figure
  \ref{fig:bpt}, are shown in gray.  Similar nuclear and extended line
  ratios are seen in all three panels, suggesting that none of the
  morphology classes have dominant nuclear AGN populations.
\label{fig:bpt_inout}}
\end{figure*}

\begin{figure*}[ht] 
\epsscale{1.15}
{\plotone{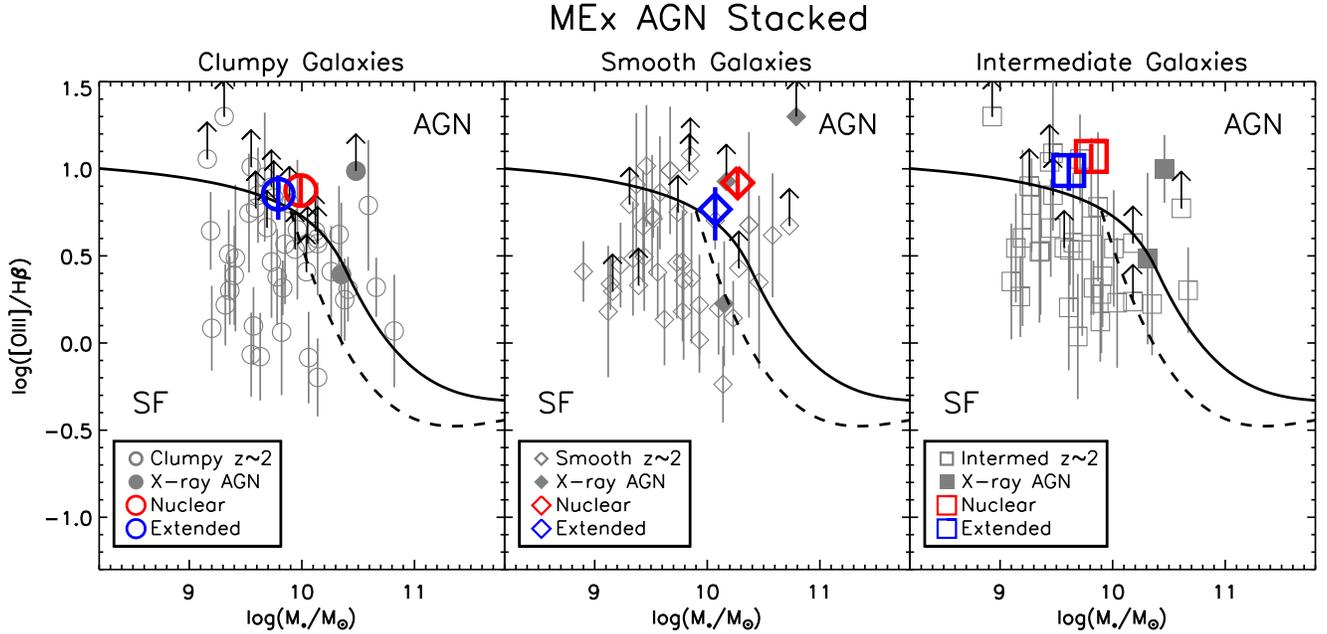}}
\figcaption{The nuclear and extended line ratios of the galaxies
  identified as AGN on the MEx diagram, stacked in each of the clumpy,
  smooth, and intermediate morphology classes.  As in Figure
  \ref{fig:bpt_inout}, we place the line ratios in the MEx diagram
  using the median $M_*$ of each stack (with a small artificial offset
  between the nuclear and extended points).  Smooth and intermediate
  MEx-classified AGNs have significantly higher nuclear ratios,
  suggesting that they are indeed likely to host AGN in their centers.
  However clumpy MEx AGNs have only marginally higher line ratios in
  their nuclear region, and much of their AGN-like $\OIII/\Hb$ ratios
  comes from extended phenomena like shocks or high-ionization star
  formation.  The spatially resolved line ratios of stacked MEx AGN in
  each morphology class suggest that clumpy galaxies may actually be
  less likely to host nuclear AGNs than smoother galaxies.
\label{fig:bpt_inoutagn}}
\end{figure*}

\begin{deluxetable}{lrr}
  \tablecolumns{3}
  \tablecaption{Stacked Spectra Nuclear and Extended $\log(\OIII/\Hb)$
    \label{tbl:stackratios}}
  \tablehead{
    \colhead{Stack} & 
    \colhead{Nuclear} & 
    \colhead{Extended} }
  \startdata
    All Clumpy & $3.5\pm0.8$ & $4.5\pm1.2$ \\
    All Smooth & $3.2\pm0.4$ & $2.6\pm1.2$ \\
    All Intermediate & $3.9\pm1.1$ & $3.3\pm1.3$ \\
    MEx AGN Clumpy & $7.5\pm1.5$ & $7.1\pm2.0$ \\
    MEx AGN Smooth & $8.3\pm1.4$ & $5.8\pm1.3$ \\
    MEx AGN Intermediate & $11.8\pm2.1$ & $9.8\pm2.3$ \\
    \OIII-centered Clumpy AGNs & $5.5\pm1.2$ & $10.5\pm1.8$
  \enddata
\end{deluxetable}

The \OIII\ and \Hb\ emission lines are measured from each nuclear and
extended spectrum following the same method as in Section 2.2: three
Gaussians were simultaneously fit to the continuum-subtracted spectra,
with the \OIII$\lambda$5007 flux given as 3/4 of the total
\OIII$\lambda$4959+5007 emission.  Figure \ref{fig:bpt_inout} shows
the nuclear and extended $\OIII/\Hb$ ratios for the stacks of all
clumpy, smooth, and intermediate galaxies in the MEx diagram (using
the median stellar mass of each sample with small offsets between the
points).  Figure \ref{fig:bpt_inoutagn} shows the nuclear and extended
line ratios using the stacked spectra of MEx AGN only.  The nuclear
and extended line ratio measurements for each stack are also presented
in Table \ref{tbl:stackratios}.

In the stacks of all galaxies in Figure \ref{fig:bpt_inout}, the
nuclear and extended spectra are nearly indistinguishable.  This
indicates that there is not a dominant AGN population in any of the
morphology classes.  Nuclear AGNs begin to marginally emerge in Figure
\ref{fig:bpt_inoutagn} in the stacked spectra of MEx AGNs among smooth
galaxies.  This implies that AGN classification using the updated
\citet{jun14} MEx method is effective for smooth galaxies at $z \sim
2$, with their stacked spectra indicating nuclear AGN.  The clumpy
galaxies classified as MEx AGN, however, have essentially no
difference in nuclear and extended $\OIII/\Hb$ ratios.  The high
$\OIII/\Hb$ ratios observed in the integrated spectra of clumpy
galaxies are probably caused by extended phenomena like shocks or
high-ionization star formation, both of which are likely in dense star
forming clumps \citep{kew13,new14,rich14}.  Spatial line ratios
indicate that clumpy galaxies may actually be \textit{less} likely to
host nuclear AGNs than smooth or intermediate galaxies.  We further
investigate the possibility of off-nuclear AGNs in Section 4.1.

\subsection{X-ray AGN}

X-rays tend to be the most efficient and least contaminated indicator
of AGN activity, and so we also use the X-ray data to quantify the AGN
fraction among clumpy, smooth, and intermediate morphology galaxies.
All galaxies are matched to the \citet{xue11} 4~Ms {\it Chandra}
catalog.  In total, 5/44 clumpy, 4/41 smooth, and 2/35 intermediate
galaxies have X-ray counterparts.  X-ray detection is not sufficient
for AGN classification, however, as luminous starburst
(SFR$\sim$300~M$_\odot$/yr) galaxies can have enough X-ray binaries to
exceed an integrated luminosity of $L_X \sim 10^{42}$
\citep{leh10,min14}.  \citet{xue11} classify the GOODS-S X-ray sources
into galaxies consistent with pure star formation or systems likely to
be AGN on the basis of their spectral shape, X-ray to optical flux
ratio, and X-ray luminosity.  These classifications reveal that 2/48
clumpy, 3/41 smooth, and 2/35 intermediate galaxies are likely to be
X-ray AGNs.  The detection fractions and full-band luminosities of the
X-ray AGNs are broadly consistent across the morphology categories,
ranging from $10^{42}-10^{44}$~erg~s$^{-1}$.  X-ray data for the
detected galaxies, as well as for the stacked non-detections
(discussed below), are presented in Table \ref{tbl:xray}.

\begin{deluxetable*}{crrrrc}
  \tablecolumns{6}
  \tablecaption{X-ray Data: Detections and Stacked Non-detections\label{tbl:xray}}
  \tablehead{
    \colhead{Galaxy ID} & 
    \colhead{Redshift} & 
    \colhead{$f$(0.5-2~keV)} &
    \colhead{$f$(2-8~keV)} &
    \colhead{$f$(0.5-8~keV)} &
    \colhead{Classification} \\
    \colhead{} & 
    \colhead{} & 
    \colhead{(erg~s$^{-1}$~cm$^{-2}$)} &
    \colhead{(erg~s$^{-1}$~cm$^{-2}$)} &
    \colhead{(erg~s$^{-1}$~cm$^{-2}$)} &
    \colhead{} }
  \startdata
8206  & 1.99 & $2.24 \times 10^{-17}$ & $<1.02 \times 10^{-16}$ & $<7.13 \times 10^{-17}$ & Galaxy \\
8409  & 1.82 & $2.26 \times 10^{-17}$ & $<1.21 \times 10^{-16}$ & $ 5.18 \times 10^{-17}$ & AGN \\
9258  & 2.02 & $1.64 \times 10^{-17}$ & $<1.05 \times 10^{-16}$ & $<6.58 \times 10^{-17}$ & Galaxy \\
9474  & 2.12 & $2.59 \times 10^{-17}$ & $<1.39 \times 10^{-16}$ & $ 7.37 \times 10^{-17}$ & Galaxy \\
10493 & 1.83 & $8.00 \times 10^{-17}$ &  $2.11 \times 10^{-16}$ & $ 2.89 \times 10^{-16}$ & AGN \\
6278  & 1.54 & $4.70 \times 10^{-17}$ &  $6.58 \times 10^{-16}$ & $ 6.75 \times 10^{-16}$ & AGN \\
9493  & 1.47 & $2.72 \times 10^{-17}$ & $<1.14 \times 10^{-16}$ & $ 6.64 \times 10^{-17}$ & Galaxy \\
10650 & 1.31 & $1.28 \times 10^{-16}$ &  $1.33 \times 10^{-16}$ & $ 2.56 \times 10^{-16}$ & AGN \\
16985 & 1.73 & $2.99 \times 10^{-16}$ &  $3.95 \times 10^{-15}$ & $ 4.03 \times 10^{-15}$ & AGN \\
9245  & 2.07 & $1.59 \times 10^{-17}$ &  $9.04 \times 10^{-17}$ & $ 8.01 \times 10^{-17}$ & AGN \\
 18315 & 2.32 & $1.57 \times 10^{-16}$ &  $6.75 \times 10^{-16}$ & $ 8.18 \times 10^{-16}$ & AGN \\
\hline
Stacked Clumpy & -- & $4.3\pm0.9 \times 10^{-18}$ & $<1.1 \times 10^{-17}$ & $9.2\pm2.8 \times 10^{-18}$ \\
Stacked Smooth & -- & $4.0\pm0.9 \times 10^{-18}$ & $<1.6 \times 10^{-17}$ & $8.8\pm2.9 \times 10^{-18}$ \\
Stacked Intermediate & -- & $<4.5 \times 10^{-18}$ & $<9.3 \times
10^{-18}$ & $<9.1 \times 10^{-18}$ \\
\hline
Stacked Clumpy MEx AGN & -- & $5.2\pm1.7 \times 10^{-18}$ & $<2.4 \times 10^{-17}$ & $13.9\pm5.1 \times 10^{-18}$ \\
Stacked Smooth MEx AGN & -- & $<5.2 \times 10^{-18}$ & $<2.6 \times 10^{-17}$ & $<17.7 \times 10^{-18}$ \\
Stacked Intermediate MEx AGN & -- & $<4.6 \times 10^{-18}$ & $<3.3 \times 10^{-17}$ & $<15.7 \times 10^{-18}$
  \enddata
\tablecomments{X-ray fluxes and AGN/galaxy classifications from
  \citet{xue11}.  X-ray stacking following the method in \citet{luo11}.}
\end{deluxetable*}

\begin{figure}[t]
\epsscale{1.15}
{\plotone{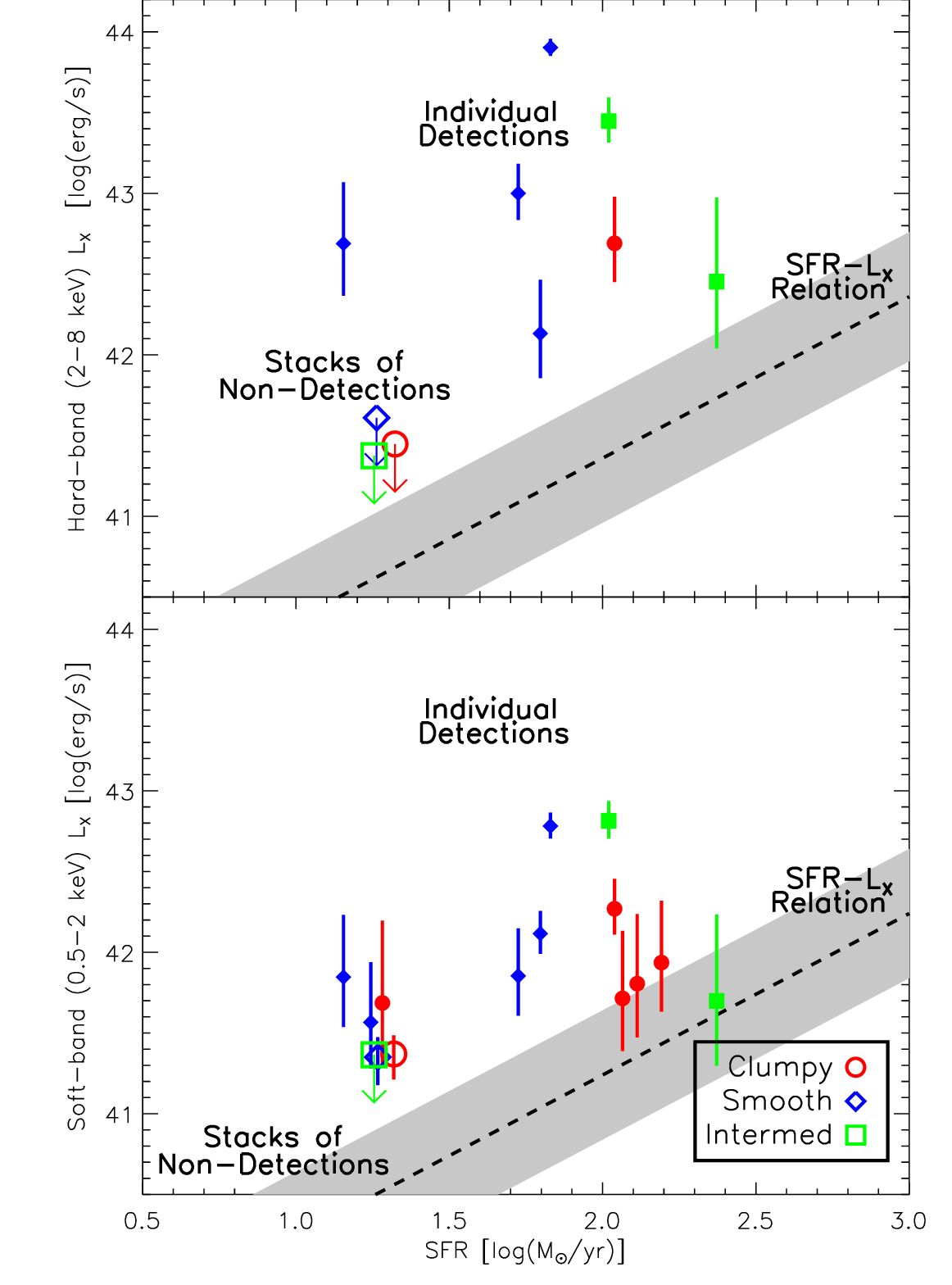}}
\figcaption{Hard-band (top panel) and soft-band (bottom panel) X-ray
  luminosities versus star formation rates, for both individually
  detected galaxies (solid points) and stacked non-detections (open
  points) among the three morphology types.  The dashed lines indicate
  the $L_X-SFR$ relation for star-forming galaxies without AGN from
  \citet{min14}, and the gray shaded region indicates the relation's
  1$\sigma$ errors.  Most individually detected sources have
  sufficiently high (hard) X-ray detections to be AGN, although the
  stacked data suggest only weak AGN emission in the X-ray
  non-detections.  In both detected sources and stacked data, there
  are no significant differences in X-ray AGN likelihood data between
  clumpy, smooth, and intermediate morphology galaxies.
\label{fig:sfrxray}}
\end{figure}

Figure \ref{fig:sfrxray} presents the $L_X-SFR$ relationship for X-ray
detected sources, for both hard-band (2--8~keV) and soft-band
(0.5--2~keV) luminosities.  Also shown is the $L_X-SFR$ relation for
star-forming galaxies without AGN, $\log(L_X)=[39.6\pm0.4]+\log(SFR)$
\citep{min14}, with $L_X$ the full-band luminosity in erg/s and $SFR$
in M$_\odot$/yr.  This relation is translated to the soft and hard
bands by assuming a spectral index of $\Gamma=1.8$: the resultant
relation is consistent with the hard-band $L_X-SFR/M_*$ relation of
\citet{leh10}.  A few X-ray sources which are undetected in the hard
band lie near the soft-band $L_X-SFR$ relation and might not host AGN,
but most have sufficient X-ray luminosities to be robustly considered
as having X-ray emission powered by an AGN.

We also performed X-ray stacking of the three samples in the full
band, soft band (0.5--2 keV), and hard band using the 4~Ms CDF-S data
\citep{xue11}.  X-ray detected galaxies and galaxies near any detected
X-ray source (within twice the soft-band 90\% encircled-energy
aperture radius of the X-ray source) were excluded from the stacking.
The final clumpy, smooth, and intermediate samples used in the
stacking contain 40, 38, and 33 galaxies, respectively.  We adopted
the same stacking procedure as described in Section 3.1 of
\citet{luo11}.  Briefly, we extracted total (source plus background)
counts for each galaxy within a 3\arcsec-diameter circular aperture
centered on its optical position.  The corresponding background counts
within this aperture were determined with a Monte Carlo approach which
randomly (avoiding known X-ray sources) places 1000 apertures within a
1\arcmin-radius circle of the optical position to measure the mean
background.  The total counts ($S$) and background counts ($B$) for
the stacked sample were derived by summing the counts for individual
sources.  The net source counts are then given by $(S-B)$, and the S/N
is calculated as $(S-B)/\sqrt{B}$.  A stacked signal is considered
undetected if it has a binomial no-source probability of $P_{\rm
  b}>0.01$ (equivalent to ${\rm S/N}\lesssim2.6\sigma$): for these
sources we derive 90\% confidence upper limits using the Bayesian
method of \citet{kra91}.  Aperture corrections were applied when
converting the source count rates to fluxes and luminosities.

The stacked luminosities are shown with the median SFR of each subset
in Figure \ref{fig:sfrxray}.  The small number of objects in each
morphology subset means that none are well-detected in the hard band,
and only the clumpy and smooth stacks are detected in the soft-band.
Due to these poor hard band detections, the hardness ratios are
unconstrained in all three stacks.  The stacked X-ray luminosities
(and upper limits) for all three samples lie only marginally above the
$L_X-SFR$ line for SF galaxies, indicating a weak or sub-dominant AGN
contribution in the X-ray undetected galaxies.  This is consistent
with the location of most galaxies in the star-forming region of the
MEx diagram (Figure \ref{fig:bpt}), and with the lack of nuclear AGN
in the resolved line ratios among all galaxies (Figure
\ref{fig:bpt_inout}).  We also stack the MEx AGNs in each morphology
class, but the numbers of objects are too small to make any meaningful
conclusions.

Clumpy, smooth, and intermediate galaxies all have similar detected
X-ray AGN fractions ($\sim$5\%) and stacked X-ray luminosities for
undetected sources ($\log(L_X) \sim 41.4$).  In agreement with the
spatially resolved line ratios, the X-ray data show essentially no
differences for clumpy, smooth, and intermediate morphology galaxies
at $z \sim 2$.

\section{Discussion}

Both spatially resolved line ratios and X-ray data (detections and
stacked non-detections) indicate that there is little or no difference
in AGN fraction between clumpy galaxies and those with smoother
morphologies.  While integrated line ratios in the MEx method suggest
a slightly higher AGN fraction among clumpy galaxies, the lack of a
nuclear AGN signature suggests that this is due to extended phenomena
like dense star-forming clumps rather than black hole growth.
Galaxies at $1.3<z<2.4$ in each morphology category have similar AGN
fractions of $\sim$5--30\%, depending on whether X-ray data or
spatially resolved line ratios are used for AGN selection.

We discuss several implications of these results below.

\subsection{Off-Nuclear AGN in Clumpy Galaxies?}

In Section 3.2 we argued that the high $\OIII/\Hb$ ratio in the
extended regions of clumpy galaxies indicates shocks or
high-ionization \HII\ regions rather than AGN activity \citep[see
also][]{rich11,new14,rich14}.  However, it is instead possible that
some clumpy galaxies host off-nuclear AGN
\citep[e.g.][]{schaw11,schaw14}.  We test for the presence of
off-nuclear AGN by creating a 2D stack of clumpy galaxies classified
as MEx-AGNs, centering on the $\OIII$ emission lines rather than the
continuum trace.  The brightest knot of $\OIII$ emission would be most
likely to correspond to any off-nuclear AGN\footnote{Our assumption
  that the brightest $\OIII$ knot corresponds to an AGN is inaccurate
  if there is large-scale dust asymmetrically obscuring an AGN
  narrow-line region.}.  The ``nuclear'' region of the
$\OIII$-centered stack corresponds to the brightest $\OIII$ knots
rather than the galaxy center, and should therefore be where
off-nuclear AGNs are most likely to reside.

\begin{figure}[t] 
\epsscale{1.15}
{\plotone{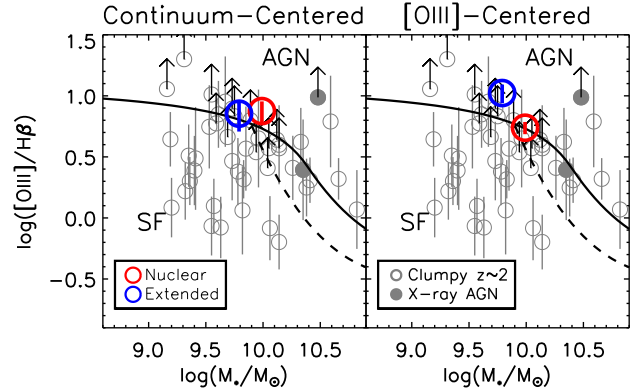}}
\figcaption{Nuclear and extended line ratios of clumpy galaxies
  classified as MEx AGNs.  The left panel uses a stacked spectrum from
  galaxies centered on their continuum trace, while the right panel
  tests for the presence of off-nuclear AGNs with a stacked spectrum
  constructed by centering each galaxy on the brightest knot of
  $\OIII$ emission.  There is no evidence for a large population of
  off-nuclear AGNs in clumpy galaxies, which would have manifested as
  a higher ``nuclear'' line ratio in the $\OIII$-centered stack.
\label{fig:bpt_linectr}}
\end{figure}

Figure \ref{fig:bpt_linectr} compares the nuclear and extended
$\OIII/\Hb$ ratios for the continuum-centered stack with the
$\OIII$-centered stack of clumpy galaxies.  The right panel suggests
that the brightest $\OIII$ knots are somewhat {\textit less} likely to
host AGNs than the galaxy centers.  The lack of widespread off-nuclear
AGNs is also supported by the X-ray data, which indicate that clumpy
galaxies do not host a greater number of X-ray AGNs (whether nuclear
or off-nuclear) than smoother mass-matched galaxies.  We conclude that
there is unlikely to be a large population of off-nuclear AGNs in
clumpy galaxies misclassified by the spatially resolved line ratio
method.

\subsection{Comparison to Lower Redshift}

At first glance, our $z \sim 1.85$ result is in apparent contrast to
the preference for AGN in clumpy galaxies at $z \sim 0.7$ observed by
\citet{bou12}.  Our clumpy galaxy sample has very similar images and
morphologies to theirs, as seen in a comparison of Figure
\ref{fig:images} with Figure 4 of \citet{bou12}.  And although we
study a much larger sample (three times more clumpy galaxies),
\citet{bou12} performed a comprehensive bootstrap analysis which
indicates that sample size is very unlikely to lead to the difference.
Our studies differ greatly, however, in the comparison samples of
smooth galaxies.

\begin{figure}[t] 
\epsscale{1.15}
{\plotone{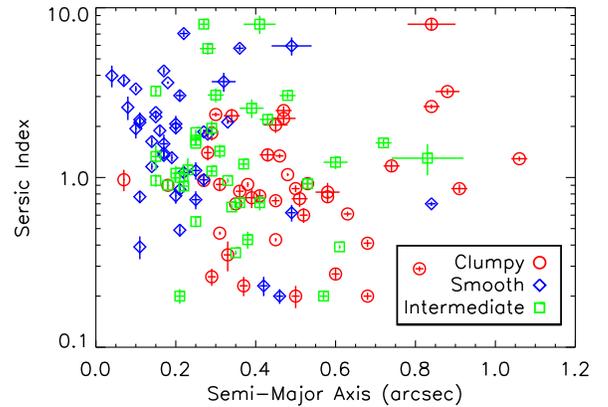}}
\figcaption{Sersic index versus size (semi-major axis) for the clumpy,
  smooth, and intermediate galaxies.  The smooth and intermediate
  galaxies are generally smaller than the clumpy galaxies, and tend to
  have more prominent bulges (although $>$70\% are still
  disk-dominated with $n<2.5$).
\label{fig:sersic}}
\end{figure}

\citet{bou12} required that both their clumpy and smooth galaxy
samples be extended disks.  Meanwhile we impose no requirements on
morphology beyond the clumpiness classification, and Figure
\ref{fig:sersic} demonstrates that our $z \sim 2$ clumpy and smooth
galaxies are quite different in size and \citet{sersic} index.  These
quantities were measured using GALFIT \citep{galfit} and presented by
\citet{vdw12}.  Clumpy galaxies are much more likely to be extended
disks, with a median semi-major axis of $0\farcs46$ and 41/44 galaxies
having disk-dominated S\'{e}rsic indices ($n<2.5$).  The smooth
galaxies are still mostly disk-dominated, as 30/41 have $n<2.5$, but
they are significantly smaller (with a median semi-major axis of
$0\farcs20$) and have higher S\'{e}rsic indices.


It is interesting that so few smooth extended disks are observed in
CANDELS and 3D-HST.  The data are sensitive to the $z \sim 2$
star-forming main sequence at $M_*/M_\odot>10^{9.8}$ (Figure
\ref{fig:sfrmass}), and so if smooth extended disks exist they must be
low-mass and/or weakly star-forming.  \citet{kas12} and \citet{ee14}
argue that ordered disks are rare in $z>1$ galaxies of any mass,
suggesting that the lack of smooth extended disks at $z \sim 2$ is a
consequence of galaxy evolution rather than selection effects.
Regardless, the lack of smooth extended disks at $z \sim 2$ means we
cannot perform the same comparison between clumpy and smooth extended
disks as \citet{bou12} at $z \sim 0.7$.  Instead we conclude that
clumpy extended galaxies have the same AGN fraction as more compact
and higher-Sersic smooth galaxies matched in stellar mass.


\subsection{Fueling AGNs at $z \sim 2$}

Our results demonstrate that clumpy galaxies at $z \sim 2$ are no more
likely to host AGNs than are smooth galaxies of the same stellar mass
and similar star formation rate.  What does this indicate about AGN
fueling modes at $z \sim 2$?

\citet{dek14} argue that compact star-forming galaxies (also called
blue nuggets) might actually be constructed by a history of violent
disk instabilities.  Our smooth galaxies, which tend to be more
compact and bulgy than the clumpy extended disks, have sizes and
Sersic indices consistent with this category \citep[see
also][]{cev14}.  The observed similarity in AGN fraction between $z
\sim 2$ clumpy and smooth galaxies would then indicate that nuclear
inflow onto the AGN must persist until after the galaxy becomes stable
and the star-forming clumps are no longer visible.  In this scenario
the violent disk instabilities are responsible for building a gas
reservoir in the galaxy's center.  Since the actual nuclear inflow
onto the AGN persists into the smooth compact galaxy phase, it must
not depend on violent instabilities in the larger disk.  So while the
gas reservoir is deposited by violent disk instabilities, the AGN
fueling within $\sim$1~kpc must instead be driven by secular
processes.

On the other hand, the disk-dominated nature of the smooth galaxies
might imply that they are unlikely to have experienced violent mixing
in the past from extreme disk instabilities (or major mergers).  It is
also not clear that the smooth galaxies are older than the clumpy
galaxies, given their similar position in the SFR-M* plane (excepting
the most massive galaxies: see Figure \ref{fig:sfrmass}).  This might
imply that even the large-scale fueling occurs via secular processes,
perhaps due to stochastic turbulence from the high gas fractions which
are common in $z \sim 2$ galaxies \citep{tac10}.  Indeed,
\citet{hop14} predict that this turbulence-driven stochastic fueling
is the dominant mode of SMBH growth for $z \sim 2$ galaxies with
$M_{BH}<10^{7}M_\odot$ (or $M_*<10^{10}M_\odot$): the same mass range
occupied by much of our sample.  Assuming that the smooth and clumpy
morphologies in our sample trace different fueling modes, then our
data best support the scenario of \citet{bel13}, where AGN growth
depends not on fueling mode, but on gas fraction.

\section{Summary}

We use spatially resolved line ratios and X-ray data to demonstrate
that $z \sim 1.85$ clumpy galaxies are no more likely to host AGNs
than are smoother galaxies at the same redshift.  While integrated
line ratios indicate a higher AGN fraction in clumpy galaxies, the
spatially resolved emission lines show that this is likely due to
extended phenomena (like shocks or high-ionization $\HII$ regions)
rather than nuclear (or off-nuclear) AGNs.  While the smooth galaxies
have the same masses and nearly the same star formation rates, they
are somewhat smaller and not quite as disk-dominated as the clumpy
galaxies.  Whatever process drives AGN fueling in the smooth galaxies
is just as efficient as are violent disk instabilities in clumpy
galaxies at $z \sim 2$.

\acknowledgements

We thank Frederic Bournaud for valuable discussion which significantly
improved this work.  Support was provided by NASA through Hubble
Fellowship grant \#51330 awarded by the Space Telescope Science
Institute, which is operated by the Association of Universities for
Research in Astronomy, Inc., for NASA under contract NAS 5-26555.  JRT
and the authors from UCSC also acknowledge support form NASA HST
grants GO-12060.10-A and AR-12822.03, Chandra grant G08-9129A, and NSF
grant AST-0808133.  SJ acknowledges financial support from the
E.C. through an ERC grant StG-257720.  This work made use of the
Rainbow Cosmological Surveys Database, which is operated by the
Universidad Complutense de Madrid (UCM), partnered with the University
of California Observatories at Santa Cruz (UCO/Lick, UCSC).

\appendix

\section{Galaxy Images}

\begin{figure*}[t] 
\begin{center}
\epsscale{1.1}
{\plotone{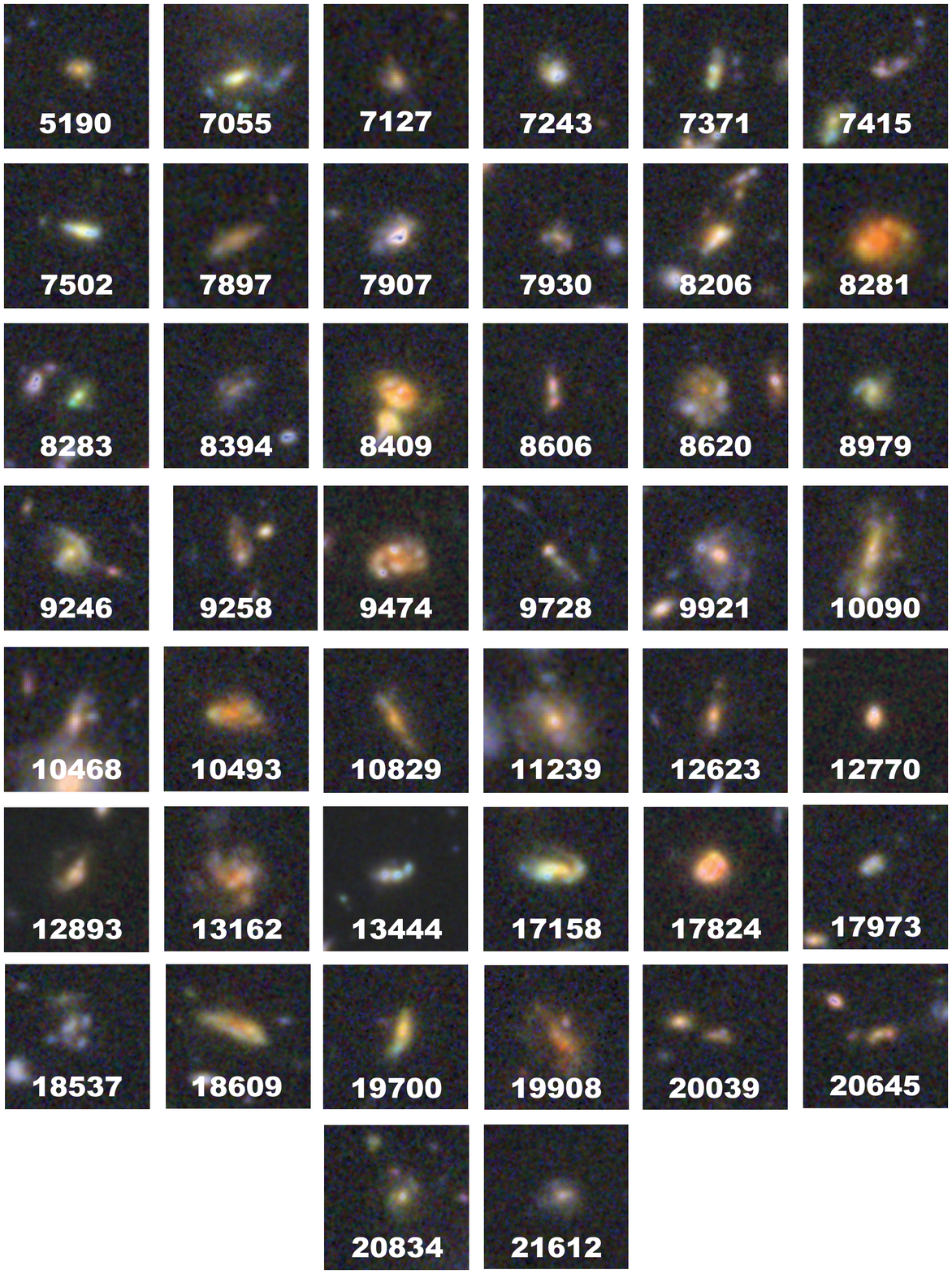}}
\end{center}
\figcaption{Images of the 44 clumpy galaxies.  Each thumbnail is
  5\arcsec\ on a side, and is a color-composite of the {\it HST} ACS
  $i$ and WFC3 $JH$ bands.
\label{fig:clumpyimages}}
\end{figure*}

\begin{figure*}[t] 
\begin{center}
\epsscale{1.1}
{\plotone{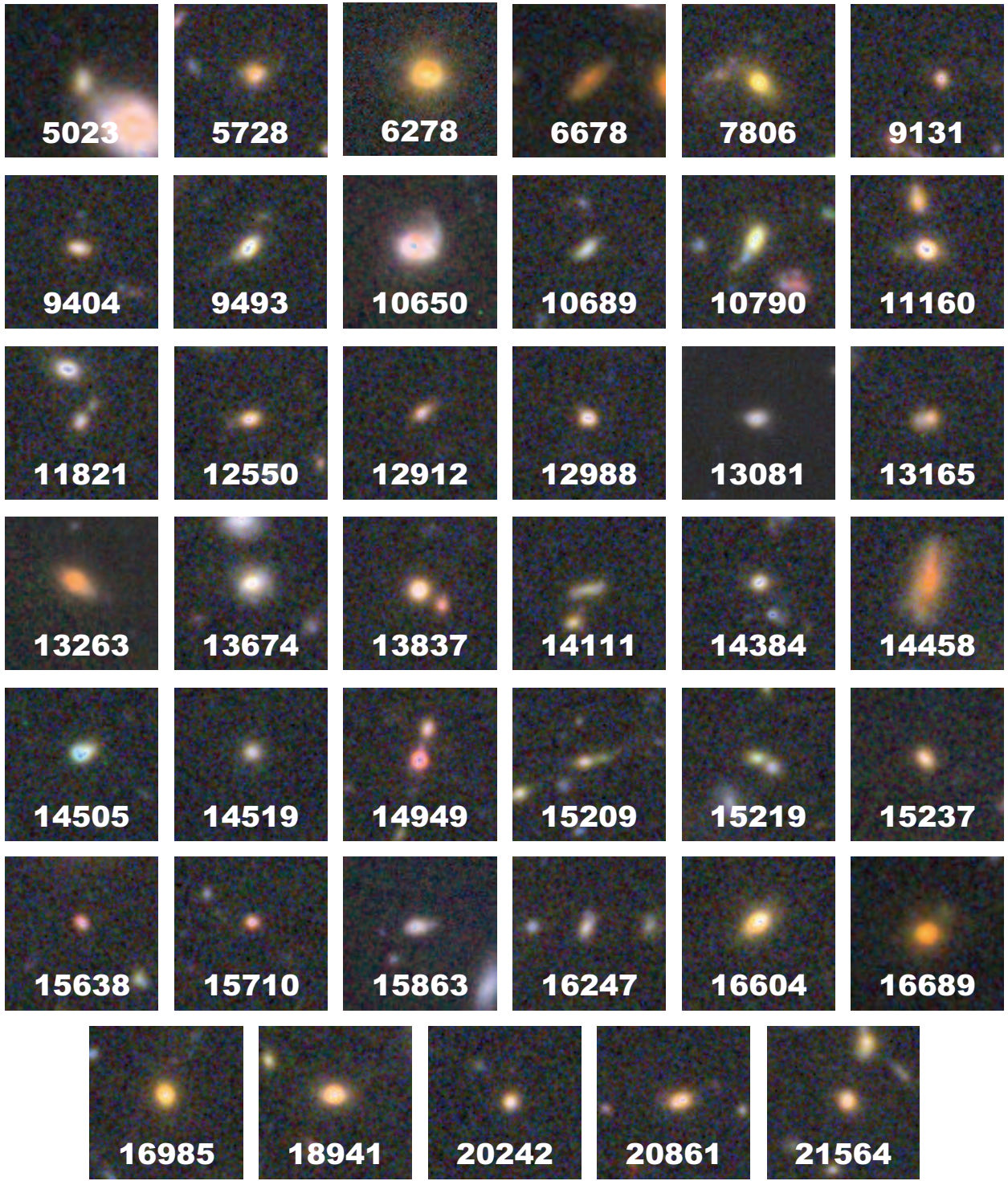}}
\end{center}
\figcaption{Color-composite $iJH$ $5\arcsec \times 5\arcsec$ images of
  the 41 smooth galaxies.
\label{fig:smoothimages}}
\end{figure*}

\begin{figure*}[t] 
\begin{center}
\epsscale{1.1}
{\plotone{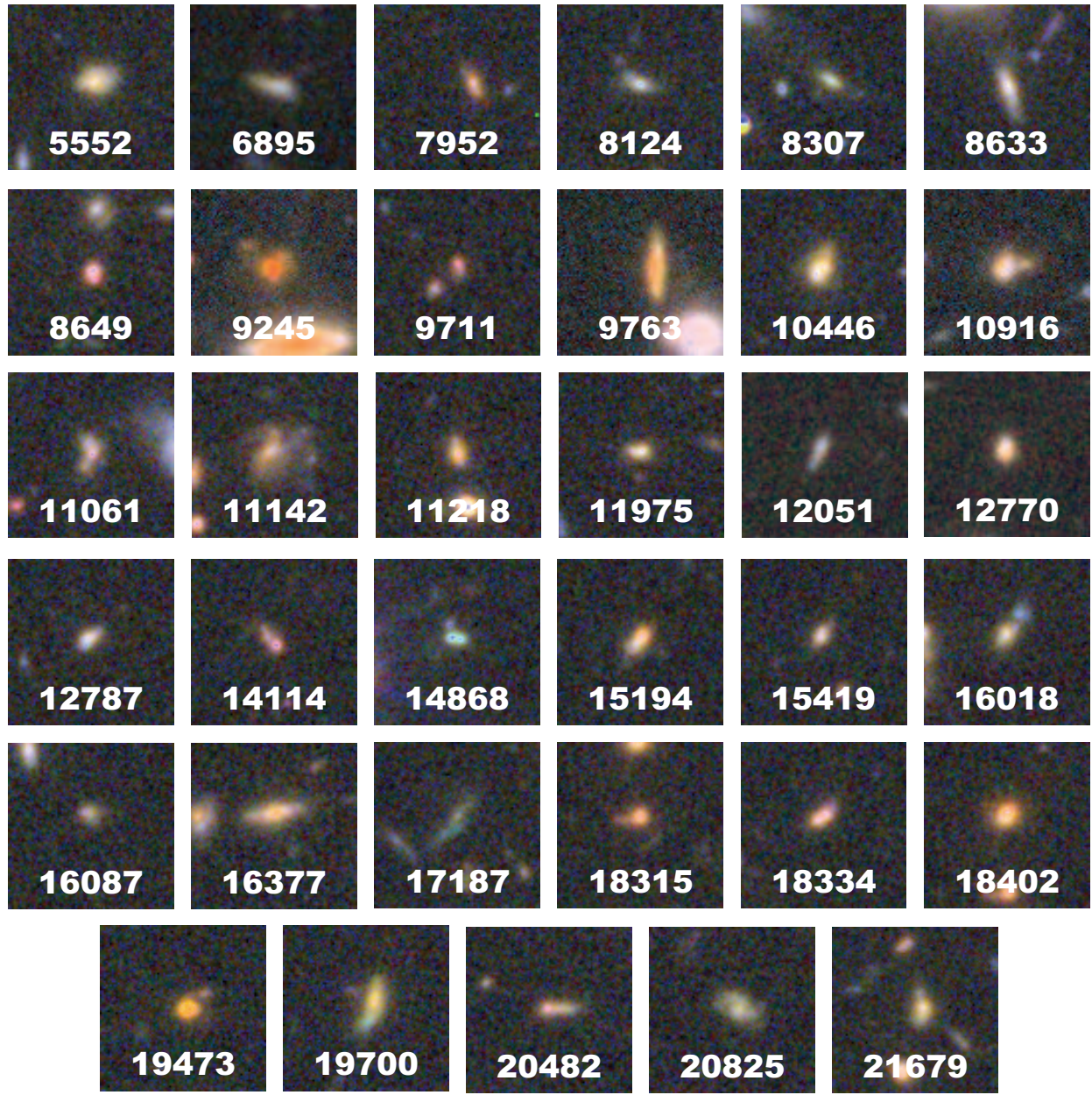}}
\end{center}
\figcaption{Color-composite $iJH$ $5\arcsec \times 5\arcsec$ images of
  the 35 intermediate-morphology galaxies.
\label{fig:intimages}}
\end{figure*}

Color-composite {\it HST} $iJH$ images of all clumpy, smooth, and
intermediate galaxies are shown in Figures \ref{fig:clumpyimages},
\ref{fig:smoothimages}, and \ref{fig:intimages}.

\end{document}